\documentclass[a4paper,11pt]{article}
\usepackage{graphicx,epsfig,picins}
\usepackage{fancyhdr,fancybox}
\usepackage{indentfirst}
\usepackage{titlesec}
\usepackage{verbatim}
\usepackage[sort&compress,numbers]{natbib}
\usepackage{array,dcolumn,tabularx}
\usepackage{setspace}
\usepackage{geometry}
\usepackage{extarrows,chemarrow,xypic} 
\usepackage[small]{caption2}

\usepackage{times}     

\usepackage{latexsym}
\usepackage{amsmath}                 
\usepackage{amssymb}                 
\usepackage{amsbsy}
\usepackage{amsthm}
\usepackage{amsfonts}
\usepackage{mathrsfs}                
\usepackage{bm}                      
\usepackage{relsize}                 

\usepackage{hyperref}  

\newcommand{\paperfont}{\fontsize{11pt}{1.1\baselineskip}\selectfont}
\begin{document}

\theoremstyle{definition}
\makeatletter
\thm@headfont{\bf}
\makeatother
\newtheorem{lemma}{Lemma}

\lhead{}
\rhead{}
\lfoot{}
\rfoot{}

\renewcommand{\refname}{References}
\renewcommand{\figurename}{Figure}
\renewcommand{\tablename}{Table}
\renewcommand{\proofname}{Proof}

\title{\textbf{An allosteric model of the inositol trisphosphate receptor with nonequilibrium binding}}
\author{Chen Jia$^{1,2}$,\;\;\;Daquan Jiang$^{1,3}$,\;\;\;Minping Qian$^{1}$ \\
\footnotesize $^1$LMAM, School of Mathematical Sciences, Peking University, Beijing 100871, P.R. China\\
\footnotesize $^2$Beijing International Center for Mathematical Research, Beijing 100871, P.R. China\\
\footnotesize $^3$Center for Statistical Science, Peking University, Beijing 100871, P.R. China\\
\footnotesize Email: jiangdq@math.pku.edu.cn}
\date{}                              
\maketitle                           
\thispagestyle{empty}                

\paperfont

\begin{abstract}
The inositol trisphosphate receptor (IPR) is a crucial ion channel that regulates the Ca$^{2+}$ influx from the endoplasmic reticulum (ER) to the cytoplasm. A thorough study of the IPR channel contributes to a better understanding of calcium oscillations and waves. It has long been observed that the IPR channel is a typical biological system which performs adaptation. However, recent advances on the physical essence of adaptation show that adaptation systems with a negative feedback mechanism, such as the IPR channel, must break detailed balance and always operate out of equilibrium with energy dissipation. Almost all previous IPR models are equilibrium models assuming detailed balance and thus violate the physical essence of adaptation. In this article, we constructed a nonequilibrium allosteric model of single IPR channels based on the patch-clamp experimental data obtained from the IPR in the outer membranes of isolated nuclei of the \emph{Xenopus} oocyte. It turns out that our model reproduces the patch-clamp experimental data reasonably well and produces both the correct steady-state and dynamic properties of the channel. Particularly, our model successfully describes the complicated bimodal [Ca$^{2+}$] dependence of the mean open duration at high [IP$_3$], a steady-state behavior which fails to be correctly described in previous IPR models. Finally, we used the patch-clamp experimental data to validate that the IPR channel indeed breaks detailed balance and thus is a nonequilibrium system which consumes energy. \\

\noindent 
\textbf{Keywords}: inositol trisphosphate receptor, adaptation, overshoot, nonequilibrium, Monod-Wyman-Changeux model
\end{abstract}

\section*{Introduction}
Cytoplasmic free Ca$^{2+}$ concentration ([Ca$^{2+}$]) plays a central role for a vast array of cellular physiological processes, such as learning and memory, muscle contraction, saliva secretion, membrane excitability, and cell division \cite{berridge1988inositol, berridge2000versatility, foskett2007inositol}. The inflow and outflow of Ca$^{2+}$ in the cytoplasm involve the Ca$^{2+}$ flux across the plasma membrane and across the internal membrane-bound compartments such as the endoplasmic reticulum (ER). One of the most important pathways of Ca$^{2+}$ influx is through the inositol (1,4,5)-trisphosphate receptor (IPR), which is an ion channel that release Ca$^{2+}$ from the ER to the cytoplasm. Structurally, the IPR channel is a tetramer of four subunits \cite{taylor2004ip}. The gating of the IPR channels requires the binding of their primary ligands, the inositol 1,4,5-trisphosphate (IP$_3$) and Ca$^{2+}$, and other ligands such as ATP and H$^+$ \cite{foskett2007inositol}. Generally, the steady-state open probability of the IPR channel is regulated by Ca$^{2+}$ with a bell-shaped [Ca$^{2+}$] dependence: Ca$^{2+}$ at low concentrations activates the channel activity, whereas Ca$^{2+}$ at higher concentrations inhibits the channel activity \cite{foskett2007inositol}. Besides the steady-state open probability, many other steady-state properties of the IPR channel were also extensively studied using patch-clamp experiments. The release of Ca$^{2+}$ from the ER can further modulate the gating of the channels, resulting in the complex behavior of Ca$^{2+}$ oscillations and waves.

\begin{figure}[!htb]
\begin{center}
\centerline{\includegraphics[width=0.8\textwidth]{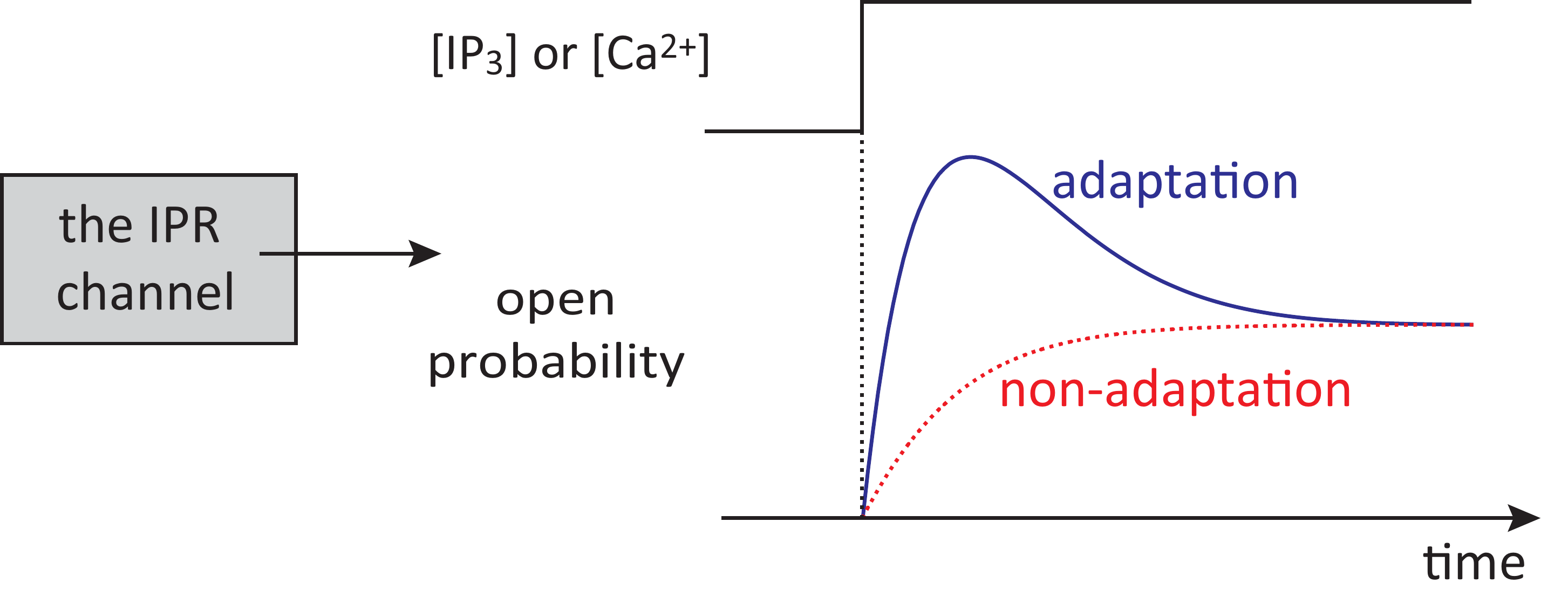}}
\caption{\textbf{Adaptation of the IPR channel}. In response to a step increase of [IP$_3$] or [Ca$^{2+}$], the open probability of the channel first rises to a peak and then declines to a lower plateau.}\label{overshoot}
\end{center}
\end{figure}

Similar to the steady-state properties, the dynamic properties of the IPR channel were also widely studied using labeled flux experiments. Recent studies show that the IPR channel responds in a time-dependent manner to a step increase of the concentration of IP$_3$ ([IP$_3$]) or Ca$^{2+}$ and performs a dynamic phenomenon called adaptation or overshoot \cite{marchant1997cooperative, marchant1998rapid, adkins1999lateral, adkins2000rapid}. An intuitive description of adaptation is depicted in Figure \ref{overshoot}, where in response to a step increase of [IP$_3$] or [Ca$^{2+}$], the open probability of the channel first rises to a peak and then declines to a lower plateau. Adaptation is one of the most important biological functions of the channel. It allows the channel to detect environmental changes more accurately, enables the channel to respond to environmental fluctuations more rapidly, and protects the channel from irreversible damages caused by unfavorable conditions.

Models of the IPR channel are essential to predict channel kinetics and understand the complex behavior of Ca$^{2+}$ oscillations and waves. Several models have been developed to describe experimental data obtained from the IPR reconstituted into artificial lipid bilayer membranes \cite{de1992single, bezprozvanny1994theoretical, kaftan1997inositol, swillens1998stochastic, moraru1999regulation}. The bell-shaped [Ca$^{2+}$] dependence of the steady-state open probability of the channel has always been a central feature in these models. However, later studies have shown that the IPR recorded in their native ER membranes behave very differently from those reconstituted into lipid bilayer membranes. Thus several models have been developed to describe experimental data obtained from the IPR channels in their native ER membrane environment \cite{sneyd2002dynamic, dawson2003kinetic, mak2003spontaneous, baran2003integrated, shuai2007kinetic, shuai2009investigation, gin2009kinetic, swaminathan2009simple, ullah2012data}.

Although adaptation of the IPR channel has been observed for more than a decade, the physics behind adaptation has remained unclear for quite a long time. In recent years, however, several research groups \cite{behar2007mathematical, franccois2008case, ma2009defining, lan2012energy} have made great efforts to study the physical essence of adaptation in biological systems. To understand how adaptation is achieved in biochemical feedback networks, Tang and coworkers \cite{ma2009defining} searched all possible three-node network topologies and found that adaptation is most likely to occur in two types of networks: the negative feedback loop and the incoherent feedforward loop. Tu and coworkers \cite{lan2012energy} further studied the physical essence of the negative feedback mechanism and found that the negative feedback loop breaks detailed balance and thus always operates out of equilibrium with energy dissipation. These two works clearly show that adaptation systems with a negative feedback mechanism are always nonequilibrium systems with energy consumption.

Interestingly, the biochemical feedback network of the IPR channel can be abstracted as a coarse-grained negative feedback loop illustrated in Figure \ref{model}(a), where IP$_3$ activates the channel activity, an increase of the channel activity facilitates the release of Ca$^{2+}$ from the ER to the cytoplasm, and Ca$^{2+}$ at high concentrations further inhibits the channel activity. Since the IPR channel is an adaptation system with a negative feedback mechanism, it should be a nonequilibrium biological system with energy consumption. Up till now, almost all previous IPR models are equilibrium models which assumes the detailed balance condition and these models apparently violate the physical essence of adaptation. The first aim of this article is to develop a nonequilibrium IPR model that produces not only the correct steady-state properties, but also the correct dynamic properties of the IPR channel.

Over the past two decades, patch-clamp experiments in the outer membranes of isolated nuclei of the \emph{Xenopus} oocyte have yielded extensive data on the gating kinetics of the IPR channels in their native ER membrane environment \cite{mak1994single, mak1997single, mak1998inositol, mak1999atp, mak2001regulation}. According to the patch-clamp experiments, the mean open duration of the IPR channel at high [IP$_3$] is regulated by Ca$^{2+}$ with a complicated bimodal dependence. Although Shuai and coworkers \cite{shuai2007kinetic} attempt to explain this phenomenon as the competition of the $A_3$ and $A_4$ openings in their model, their explanation is not so successful since their theoretical expression of the mean open duration is always a monomodal function of [Ca$^{2+}$]. So far, none of previous IPR models can produce the correct bimodal dependence of the mean open duration on [Ca$^{2+}$]. The second aim of this article is to develop an allosteric IPR model that not only reproduces the patch-clamp experimental data obtained from the nuclear IPR of \emph{Xenopus} oocytes but also produces correct dependence of the mean open duration on [Ca$^{2+}$].

In this article, we constructed a nonequilibrium allosteric model of the IPR channel using the patch-clamped experimental data obtained from the nuclear IPR of \emph{Xenopus} oocytes. Our allosteric model is composed of models at two different levels, the subunit model and the channel model. For each IPR subunit, we continued to use the model developed by Shuai and coworkers \cite{shuai2007kinetic} except that we assumed that each subunit can exist in two configurations. Inspired by the classical Monod-Wyman-Changeux allosteric model, we then constructed our model from the subunit level to the channel level. Different from previous IPR models, our subunit model is a nonequilibrium model without the assumption of detailed balance. The removal of the detailed balance condition adds an extra complexity in the model analysis. However, we overcame this difficulty successfully by using the mathematical tool of the circulation theory of Markov chains \cite{jiang2004mathematical}.

We showed that our allosteric model reproduces the patch-clamp recordings of the nuclear IPR of \emph{Xenopus} oocytes at different concentrations of IP$_3$ and Ca$^{2+}$ reasonably well. Particularly, our model successfully describes the complicated bimodal [Ca$^{2+}$] dependence of the mean open duration at high [IP$_3$], a steady-state phenomenon that fails to be correctly described in previous IPR models, and reveals that the breakdown of detailed balance in the IPR channel gives rise to this complicated bimodal behavior. Moreover, our model successfully describes the dynamic phenomenon of adaptation. By carefully checking the rate constants obtained from the data fitting, we found that two parameters in our subunit model are very close to zero. This fact shows that there is an apparent breakdown of detailed balance in the subunit model, and thus implies that the IPR channel is indeed a nonequilibrium system which consumes energy.

\section*{Model}

\subsection*{Why do we need a nonequilibrium model?}
Over the past decade, the developments of labeled flux experiments have shown that the IPR channel is a typical biological system that performs adaptation \cite{marchant1997cooperative, marchant1998rapid, adkins1999lateral, adkins2000rapid}. Although adaptation has been widely observed in various kinds of biological systems, the physical essence of adaptation has remained unclear for quite a long time.

Recently, significant progresses have been made in the study of the physical essence of adaptation \cite{behar2007mathematical, franccois2008case, ma2009defining, lan2012energy}. These results have become one of the most important developments in biophysics in recent years. Among these works, Tu and coworkers \cite{lan2012energy} studied adaptation systems with a negative feedback mechanism in great detail and found that the negative feedback loop breaks detailed balance and always operates out of equilibrium with energy dissipation. Interestingly, the biochemical feedback network of the IPR channel can be exactly abstracted as a coarse-grained negative feedback loop illustrated in Figure \ref{model}(a). This clearly shows that the IPR channel, as an adaptation system with a negative feedback mechanism, must finally approach a nonequilibrium steady state.

Here, we have used the concepts of equilibrium and nonequilibrium steady states in nonequilibrium statistical physics, where the steady state of a system is called equilibrium (nonequilibrium) if the system satisfies (breaks) the detailed balance condition, which requires that for each pair of states $i$ and $j$ of the system, the probability flux from $i$ to $j$ is always the same as that from $j$ to $i$. In the following discussion, a system that will finally approach an equilibrium (nonequilibrium) steady state is referred to as an equilibrium (nonequilibrium) system. From the viewpoint of statistical physics, an equilibrium system in the steady state is microscopic reversible and does not consume energy, whereas a nonequilibrium system is microscopic irreversible and always consumes energy. Moreover, an equilibrium system is usually a closed system which has no life, whereas a nonequilibrium system must be an open system which constantly exchanges materials and energy with its environment \cite{qian2006open,qian2007phosphorylation}.

So far, almost all previous IPR models are equilibrium models satisfying the detailed balance condition, mainly for the following two reasons. First, the authors of previous models tended to treat the IPR channel as a closed system and did not consider the fact that living systems will constantly exchange materials and energy with their environment. Second, the detailed balance condition reduces the complexities of calculations and formulations to a remarkable extent. However, we have seen that equilibrium models violate the dissipative nature of adaptation. Thus we need to develop a nonequilibrium IPR model that produces both the correct steady-state properties and the correct dynamic properties of the IPR channel.

\subsection*{The subunit model}
The structural studies show that the IPR channel is a tetramer of four subunits \cite{taylor2004ip}. As a highly allosteric protein, the IPR channel is regulated by several heterotropic ligands including its primary ligands, IP$_3$ and Ca$^{2+}$, and other ligands such as ATP, H$^+$, and interacting proteins, as well as by redox and phosphorylation status \cite{foskett2007inositol}. So far, most experiments about the IPR channel mainly focus on studying how the gating of the channel is regulated by their primary ligands, IP$_3$ and Ca$^{2+}$. However, the binding affinities of the primary ligands will be strongly influenced by the conformational state of the channel, which is in turn dependent on the binding state of all those non-primary ligands. As a result, if we abstract the binding state of all those non-primary ligands into different configurations, then we have good reasons to believe that each IPR subunit has two or more configurations, each of which corresponds to a binding state of those non-primary ligands. In this article, for simplicity, we assume that each IPR subunit can exist in two different configurations, $R$ and $T$.

\begin{figure}[!htb]
\begin{center}
\centerline{\includegraphics[width=1.0\textwidth]{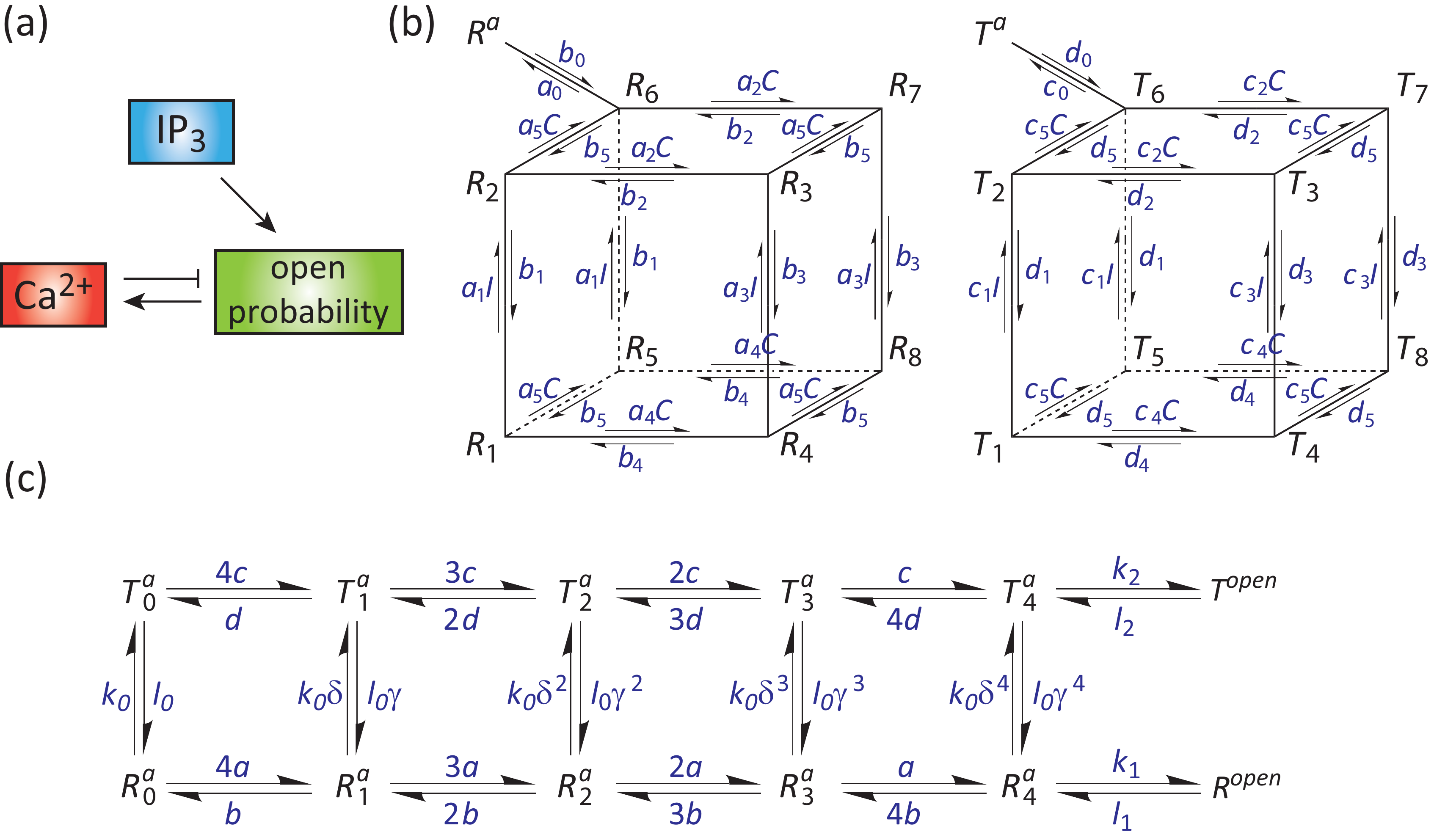}}
\caption{\textbf{(a)-(c) Models of the IPR channel.}
\textbf{(a)} A coarse-grained negative feedback loop in the IPR channel. In this feedback network, IP$_3$ activates the channel activity, an increase of the channel activity facilitates the release of Ca$^{2+}$ from the ER to the cytoplasm, and Ca$^{2+}$ at high concentrations further inhibits the channel activity.
\textbf{(b)} The subunit models. Each $R$ or $T$ subunit has an IP$_3$ binding site, together with two Ca$^{2+}$ binding sites, an activating site and an inactivating sites. The $R$ or $T$ subunit is potentiated when it is at the state $R_6$ or $T_6$ and is activated when it is at the state $R^a$ or $T^a$. We make two simplifying assumptions about the rate constants. First, we assume that the rate constants are independent of whether activating Ca$^{2+}$ is bound or not. Second, we assume that the kinetics of Ca$^{2+}$ activation are independent of IP$_3$ binding and Ca$^{2+}$ inactivation. Under these two assumptions, some rate constants are regarded as the same.
\textbf{(c)} The channel model. In the channel model, all of the four IPR subunits are in the same configuration at any time. According to the numbers of the activated and inactivated subunits, each channel has five possible states, $R^a_i$ ($i = 0,1,2,3,4$), corresponding to the configuration $R$ and five possible states, $T^a_i$ ($i = 0,1,2,3,4$), corresponding to the configuration $T$. Moreover, the state $R^a_i$ and its mirror state $T^a_i$ can convert between each other. The channel is potentiated when it is at one of its rightmost closed states, $R^a_4$ and $T^a_4$, and is open when it is at one of its open states, $R^{open}$ and $T^{open}$.}\label{model}
\end{center}
\end{figure}

For each subunit, we continue to use the model developed by Shuai, Pearson, Foskett, Mak, and Parker (abbreviated as the SPFMP model) \cite{shuai2007kinetic}. The transition diagrams of the $R$ and $T$ subunits are depicted in Figure \ref{model}(b), where we assume that the subunits in two different configurations have the same transition diagram with different rate constants. We shall explain our subunit model only for the $R$ subunit, since that for the $T$ subunit is totally the same.

Structurally, each IPR subunit is known to have at least one IP$_3$ binding site and two Ca$^{2+}$ binding sites \cite{foskett2007inositol}. Based on the experimental result that there is a bell-shaped [Ca$^{2+}$] dependence of the channel open probability, we assume that each subunit has two independent Ca$^{2+}$ binding sites: an activating binding site and an inhibitory biding site. Thus eight states, $R_1,\cdots,R_8$, are introduced to describe the kinetics of the $R$ subunit according to whether the three binding sites, an IP$_3$ binding site and two Ca$^{2+}$ binding sites, are occupied or not. We assume that each subunit is potentiated when the IP$_3$ and activating Ca$^{2+}$ binding sites are occupied, but the inhibitory Ca$^{2+}$ binding site is not occupied. Under this assumption, IP$_3$ and low concentration of Ca$^{2+}$ will promote the subunit activity, whereas high concentration of Ca$^{2+}$ will lead to inhibition. In this way, the negative feedback mechanism in Fig. \ref{model}(a) is realized in our subunit model. The subunit model further includes a conformational change whereby a subunit with the IP$_3$ and activating Ca$^{2+}$ binding sites occupied is inactivated, and must transfer to an activated state $R^a$ before it can contribute to the channel opening. The eight states, $R_1,\cdots,R_8$, and an extra activated state $R^a$ constitute a total of nine states of the $R$ subunit. In the following discussion, we collectively refer to the eight states, $R_1,\cdots,R_8$, as the inactivated state $R^i$. In this way, each $R$ subunit can be approximately considered to convert between its activated state $R^a$ and its inactivated state $R^i$.

To simplify notations, we denote [IP$_3$] and [Ca$^{2+}$] by $I$ and $C$, respectively. We see from Figure \ref{model}(b) that the kinetics of each $R$ subunit is governed by pseudo-first-order rate constants $a_1I, a_2C, a_3I, a_4C$, and $a_5C$ for the binding processes, first-order rate constants $b_1, b_2, b_3, b_4$, and $b_5$ for the unbinding process, and constant transition rates $a_0$ and $b_0$ for the transitions between states $R_6$ and $R^a$. We emphasize here that we do not require that our subunit model satisfies the detailed balance condition. Thus our subunit model is a nonequilibrium model.

\subsection*{The channel model}
We have constructed a model for each IPR subunit. The remaining question is to construct our IPR model from the subunit level to the channel level. Different from many previous IPR models which assumed that all subunits are independent \cite{de1992single, shuai2007kinetic, shuai2009investigation}, we assume that the four IPR subunits are cooperative so that all subunits are in the same configuration at any time. We make this assumption mainly for the following two reasons. First, the patch-clamp experiments show that there is a bimodal [Ca$^{2+}$] dependence of the mean open duration at high [IP$_3$]. If we assume that all subunits are independent, then we can prove mathematically that the mean open duration and the open probability must depend on [Ca$^{2+}$] in the same way. This suggests that if we assume that all subunits are independent, then the channel open probability will be regulated by Ca$^{2+}$ with a bimodal [Ca$^{2+}$] dependence, which is inconsistent with the observed bell-shaped [Ca$^{2+}$] dependence. Thus we have good reasons to believe that the four IPR subunits are strongly cooperative. Second, if we do not make any cooperative assumption, then the channel model will have a total number of $4^4 = 256$ states, which are so large that none of the rate constants can be estimated robustly from experimental data.

We have assumed that all of the four IPR subunits are in the same configuration at any time. According to the numbers of activated and inactivated subunits, each channel has five possible states, $R^a_0$, $R^a_1$, $R^a_2$, $R^a_3$, and $R^a_4$, corresponding to the configuration $R$, where $R^a_i$ ($i = 0,1,2,3,4$) represents the state that the channel has $i$ activated $R$ subunits and $4-i$ inactivated ones. Similarly, each channel has five mirror states, $T^a_0$, $T^a_1$, $T^a_2$, $T^a_3$, and $T^a_4$, corresponding to the configuration $T$. The transition diagram of our channel model is depicted in Figure \ref{model}(c), where we assume that the state $R^a_i$ and its mirror state $T^a_i$ can convert between each other. We further assume that the above ten states are all closed states. When all of the four IPR subunits are activated, that is, when the channel is at one of its rightmost closed states, $R^a_4$ and $T^a_4$, it may convert into one of its open states, $R^{open}$ and $T^{open}$. The basic idea of our channel model is similar to the classical Monod-Wyman-Changeux allosteric model, which is widely used in modeling various kinds of receptor systems in living cells \cite{jafri1998cardiac, stern1999local, tu2008modeling}.

The transitions between the states $R^a_i$ ($i=0,1,2,3,4$) are governed by rate constants $a$ and $b$, where $a$ represents the rate constant from the inactivated state $R^i$ to the activated state $R^a$, and $b$ represents that from the activated state $R^a$ to the inactivated state $R^i$. Moreover, the transitions between the closed state $R^a_4$ and the open state $R^{open}$ are governed by transition rates $k_1$ and $l_1$. In addition, the transitions between the state $R^a_i$ and its mirror state $T^a_i$ ($i=0,1,2,3,4$) are governed by transition rates $k_0\delta^i$ and $l_0\gamma^i$. The additional constants $\delta$ and $\gamma$ are introduced to make the channel model satisfy the detailed balance condition, which requires that for each cycle, the product of rate constants in the clockwise direction is equal to that in the counterclockwise direction. Then we easily see that $\delta$ and $\gamma$ must satisfy
\begin{equation}\label{constants}
\delta ad = \gamma bc.
\end{equation}
We emphasize here that we should have removed the detailed balance condition in both the subunit and channel models. However, this will make the calculations and formulations extremely complicated. Thus, for the sake of simplicity, we assume that our channel model satisfies the detailed balance condition, but we do not make the same assumption on our subunit model. Thus overall, our allosteric model of the IPR channel is a nonequilibrium model.

By fitting the patch-clamp experimental data obtained from the nuclear IPR of \emph{Xenopus} oocyte from recent published work \cite{mak2003spontaneous}, we obtain a set of optimal parameters (binding rate constants, unbinding rate constants, and constant transition rates) for our subunit and channel models. The data used to estimate the parameter values include the open probability data and the mean open duration data at high [IP$_3$] of 10 $\mu$M and low [IP$_3$] of 0.1 $\mu$M. Since the units of the measurements of the open probability and the mean open duration are different, we adopt the weighted least-square criterion to estimate the parameters, where the mean open duration data are properly prioritized. We use the parameters in the SPFMP model with random perturbations as initial parameters for the optimization. According to our simulation, all equilibrium constants, $a_i/b_i$ ($i=0,1,2,3,4,5$) and $k_i/l_i$ ($i=0,1,2$), and two additional parameters, $l_1$ and $l_2$, converge to the same final values for random initial parameters and thus can be estimated robustly. The other parameters are determined so that the predicted time scales of adaptation are consistent with the observed time scales in the label flux experiments. The optimal parameters that we have estimated are listed in Table \ref{parameters}.
\begin{table}[!htb]
\centering
\begin{tabular}{|c|c|c|}
\hline
                                  & Parameter & Value           \\ \hline
Conformational change             & $a_0$     & 5.35$\times10^{-1}$ ms$^{-1}$ \\
                                  & $b_0$     & 1.33$\times10^{-1}$ ms$^{-1}$ \\
IP$_3$ binding site               & $a_1$     & 8.97$\times10^{-6}$ $\mu$M$^{-1}$ms$^{-1}$ \\
                                  & $b_1$     & 5.19$\times10^{-3}$ ms$^{-1}$ \\
Inhibitory Ca$^{2+}$ binding site & $a_2$     & 1.28$\times10^{-3}$ $\mu$M$^{-1}$ms$^{-1}$ \\
                                  & $b_2$     & 2.24$\times10^{-2}$ ms$^{-1}$ \\
IP$_3$ binding site               & $a_3$     & 2.04 $\mu$M$^{-1}$ms$^{-1}$ \\
                                  & $b_3$     & 3.18$\times10^{-1}$ ms$^{-1}$ \\
Inhibitory Ca$^{2+}$ binding site & $a_4$     & 1.72$\times10^{-1}$ $\mu$M$^{-1}$ms$^{-1}$ \\
                                  & $b_4$     & 4.24$\times10^{-2}$ ms$^{-1}$ \\
Activating Ca$^{2+}$ binding site & $a_5$     & 1.51$\times10^{-1}$ $\mu$M$^{-1}$ms$^{-1}$ \\
                                  & $b_5$     & 7.87$\times10^{-2}$ ms$^{-1}$ \\ \hline
Conformational change             & $c_0$     & 5.43$\times10^{-1}$ ms$^{-1}$ \\
                                  & $d_0$     & 7.70$\times10^{-2}$ ms$^{-1}$ \\
IP$_3$ binding site               & $c_1$     & 5.35$\times10^{-1}$ $\mu$M$^{-1}$ ms$^{-1}$ \\
                                  & $d_1$     & 1.64$\times10^{-2}$ ms$^{-1}$ \\
Inhibitory Ca$^{2+}$ binding site & $c_2$     & 6.42$\times10^{-8}$ $\mu$M$^{-1}$ms$^{-1}$ \\
                                  & $d_2$     & 1.56$\times10^{-3}$ ms$^{-1}$ \\
IP$_3$ binding site               & $c_3$     & 1.22 $\mu$M$^{-1}$ms$^{-1}$ \\
                                  & $d_3$     & 7.00$\times10^{-3}$ ms$^{-1}$ \\
Inhibitory Ca$^{2+}$ binding site & $c_4$     & 1.69$\times10^{-1}$ $\mu$M$^{-1}$ms$^{-1}$ \\
                                  & $d_4$     & 7.40$\times10^{-1}$ ms$^{-1}$ \\
Activating Ca$^{2+}$ binding site & $c_5$     & 1.50$\times10^{-1}$ $\mu$M$^{-1}$ms$^{-1}$ \\
                                  & $d_5$     & 2.34$\times10^{-1}$ ms$^{-1}$ \\ \hline
Conformational change             & $k_0$     & 1.00 ms$^{-1}$ \\
                                  & $l_0$     & 6.57$\times10^{-1}$ ms$^{-1}$ \\
Conformational change             & $k_1$     & 2.63 ms$^{-1}$ \\
                                  & $l_1$     & 5.87$\times10^{-2}$ ms$^{-1}$ \\
Conformational change             & $k_2$     & 1.53 ms$^{-1}$ \\
                                  & $l_2$     & 3.17 ms$^{-1}$ \\ \hline
\end{tabular}
\caption{The model parameters (rate constants for binding and unbinding processes and transition rates of conformational changes) estimated by applying our allosteric model to fit the patch-clamp experimental data obtained from the nuclear IPR of \emph{Xenopus} oocytes. The optimal values of the model parameters are estimated based on the weighted least-square criterion.}\label{parameters}
\end{table}

The parameters $a$ and $b$ in the channel model represent the transition rates between the inactivated state $R^i$ and activated state $R^a$, and thus must be functions of the rate constants $a_i$ and $b_i$ ($i=0,1,2,3,4,5$) in the subunit model. A difficult point is to determine how the parameters $a$ and $b$ depend on the rate constants $a_i$ and $b_i$. In this paper, we use the probability definition of the transition rates and the circulation theory of Markov chains \cite{jiang2004mathematical} to derive the specific expressions of $a$ and $b$. To make our discussion friendly to those unfamiliar with these mathematical tools, we would like to present the results here and put the detailed derivation in \emph{Methods}, from which we obtain that
\begin{equation}\label{rateR}
a = a_0\times\frac{a_5C}{a_5C+b_5}\times\frac{Q_2}{Q_1+Q_2+Q_3+Q_4},~~~b = b_0,
\end{equation}
where
\begin{equation}\label{determinant}
\begin{split}
Q_1 &= b_1b_2a_3I + a_2b_3b_4C + b_1b_2b_4 + b_1b_3b_4; \\
Q_2 &= (a_1b_2a_3I + b_2a_3a_4C + a_1b_2b_4 + a_1b_3b_4)I; \\
Q_3 &= (a_1a_2a_3I + a_2a_3a_4C + a_1a_2b_4 + b_1a_3a_4)IC; \\
Q_4 &= (a_1a_2b_3I + a_2b_3a_4C + b_1b_2a_4 + b_1b_3a_4)C.
\end{split}
\end{equation}
Similar expressions can be obtained for the $T$ subunit. The parameters $c$ and $d$ in the channel model can be calculated as
\begin{equation}\label{rateT}
c = c_0\times\frac{c_5C}{c_5C+d_5}\times\frac{R_2}{R_1+R_2+R_3+R_4},~~~d = d_0,
\end{equation}
where $R_i$ ($i=1,2,3,4$) is obtained from $Q_i$ by changing those $a_j$ ($j=1,2,3,4$) in Equation \eqref{determinant} to $c_j$ and by changing those $b_j$ ($j=1,2,3,4$) in Equation \eqref{determinant} to $d_j$.

\section*{Results}

\subsection*{General analysis}
In this section, we shall give the theoretical expressions of four important quantities related to the gating kinetics of the IPR channel: the (steady-state) open probability $P_o$, mean closed duration $\tau_c$, mean open duration $\tau_o$, and the distribution of the open duration $p_o(t)$.

Let $p$ and $q$ denote the steady-state probabilities of the two open states, $R^{open}$ and $T^{open}$, in the channel model, respectively. Since we have assumed that the channel model satisfies the detailed balance condition, we easily see that
\begin{equation}\label{detailedbalance}
pl_1k_0\delta^4k_2 = ql_2l_0\gamma^4k_1.
\end{equation}
To simplify notations, we introduce three equilibrium constants as $K_0 = k_0/l_0$, $K_1 = k_1/l_1$, and $K_2 = k_2/l_2$. Then Equation \eqref{detailedbalance} can be rewritten as
\begin{equation}\label{dbrewritten}
pK_0K_2\delta^4 = qK_1\gamma^4.
\end{equation}
Since the sum of the steady-state probabilities of all states in the channel model equals to 1, we obtain that
\begin{equation}\label{normalization}
p + p\frac{l_1}{k_1}\left(1+\frac{b}{a}\right)^4 + q + q\frac{l_2}{k_2}\left(1+\frac{d}{c}\right)^4 = 1.
\end{equation}
We further introduce two constants as $K_R = a/b$ and $K_T = c/d$. It then follows from Equations \eqref{constants}, \eqref{dbrewritten}, and \eqref{normalization} that
\begin{equation}
\begin{split}
p &= \frac{K_1K_R^4}{K_1K_R^4+K_0K_2K_T^4+(1+K_R)^4+K_0(1+K_T)^4};\\
q &= \frac{K_0K_2K_T^4}{K_1K_R^4+K_0K_2K_T^4+(1+K_R)^4+K_0(1+K_T)^4}.
\end{split}
\end{equation}
Thus the open probability $P_o$ of the IPR channel is given by
\begin{equation}\label{openprobformula}
P_o = p+q = \frac{K_1K_R^4+K_0K_2K_T^4}{K_1K_R^4+K_0K_2K_T^4+(1+K_R)^4+K_0(1+K_T)^4}.
\end{equation}

Next, we calculate the mean open and closed durations of the IPR channel. It can be proved that the mean open duration $\tau_o$ is exactly the quotient of the steady-state open probability $P_o$ and the total probability flux between the open states and the closed states \cite{shuai2007kinetic}, where the open states are the two states, $R^{open}$ and $T^{open}$, and the closed states are the rest states in the channel model. In this way, the mean open duration $\tau_o$ of the IPR channel can be calculated as
\begin{equation}\label{opentimeformula}
\tau_o = \frac{p+q}{l_1p+l_2q} = \frac{K_1K_R^4+K_0K_2K_T^4}{l_1K_1K_R^4+l_2K_0K_2K_T^4}.
\end{equation}
Similarly, the mean closed duration is the quotient of the steady-state closed probability $1-P_{o}$ and the total probability flux between the open states and the closed states. Thus the mean closed duration $\tau_c$ of the IPR channel can be calculated as
\begin{equation}\label{closedtimeformula}
\tau_c = \frac{1-(p+q)}{l_1p+l_2q} = \frac{(1+K_R)^4+K_0(1+K_T)^4}{l_1K_1K_R^4+l_2K_0K_2K_T^4}.
\end{equation}

Finally, we consider the distribution of the open duration. Since there are only two open states, $R^{open}$ and $T^{open}$, the distribution of the open duration must have a bi-exponential distribution, whose distribution density $p_o(t)$ is given by
\begin{equation}\label{biexp}
\begin{split}
p_o(t) &= \frac{l_1p}{l_1p+l_2q}l_1e^{-l_1t} + \frac{l_2q}{l_1p+l_2q}l_2e^{-l_2t} \\
&= \frac{l_1^2K_1K_R^4e^{-l_1t}+l_2^2K_0K_2K_T^4e^{-l_2t}}{l_1K_1K_R^4+l_2K_0K_2K_T^4}.
\end{split}
\end{equation}
The distribution density $p_o(t)$ of the open duration is easily seen to be the weighted sum of two exponential distribution density with time constants $l_1$ and $l_2$, respectively. The weights $f_R = l_1p/(l_1p+l_2q)$ and $f_T = l_2q/(l_1p+l_2q)$ in the summation are just the fractions of the $R^{open}$ openings and the $T^{open}$ openings, respectively.

We have expressed the four important quantities related to the gating kinetics of the IPR channel as functions of $l_1,l_2,K_0,K_1,K_2,K_R$, and $K_T$. Based on these theoretical expressions and the parameters listed in Table \ref{parameters}, we can make model predictions about these four quantities at different [IP$_3$] and [Ca$^{2+}$].

\subsection*{Open probability}
By fitting the patch-clamp experimental data obtained from the nuclear IPR of \emph{Xenopus} oocyte, we have estimated the optimal parameters in our subunit and channel models. In the following discussion, we shall show that our allosteric model with the above parameters successfully reproduces the patch-clamp recordings. Moreover, we shall make some predictions based on our allosteric model.

\begin{figure}[!htb]
\begin{center}
\centerline{\includegraphics[width=0.8\textwidth]{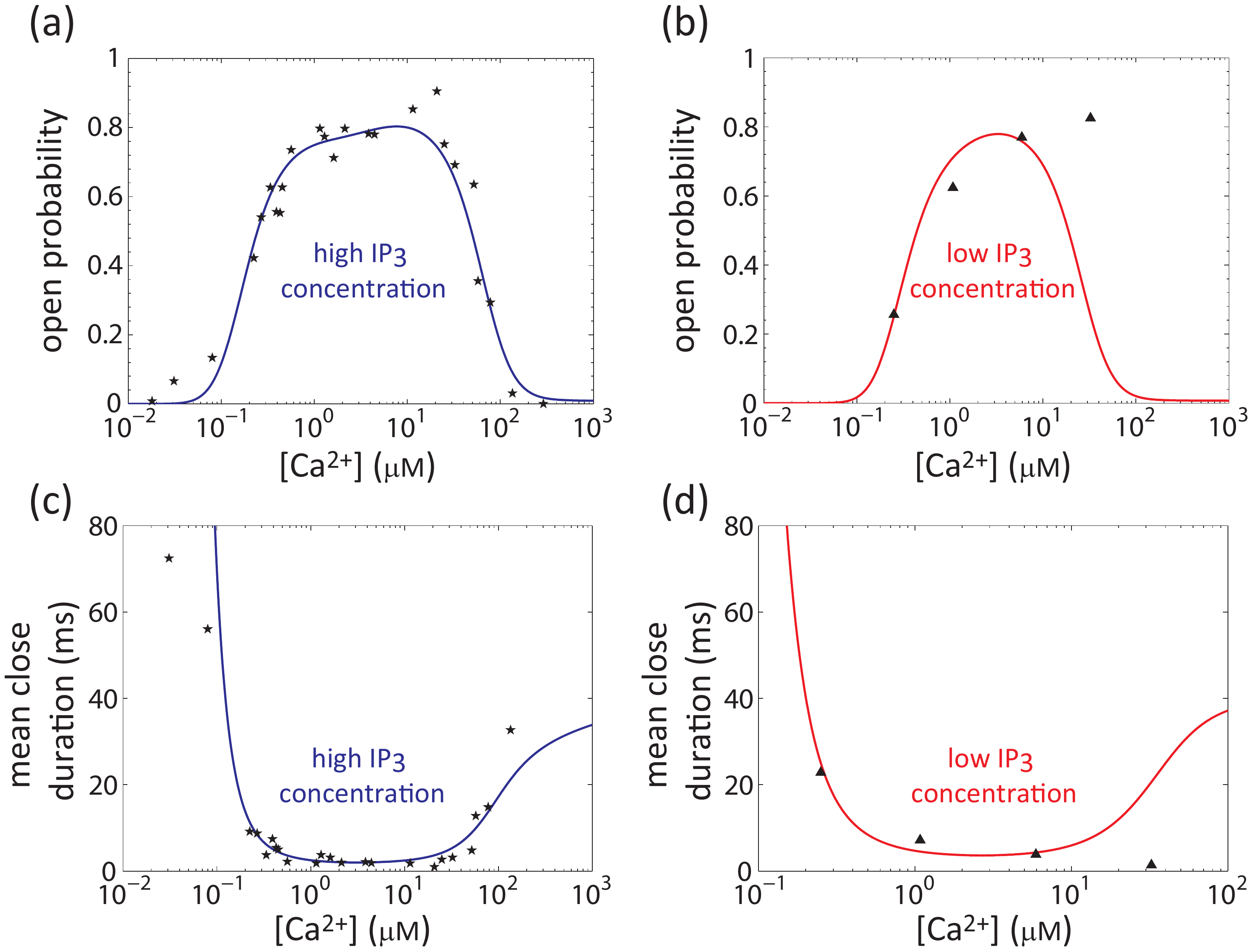}}
\caption{\textbf{(a)-(d) The [Ca$^{2+}$] dependence of the open probability and mean closed duration at different [IP$_3$].}
\textbf{(a)-(b)} The [Ca$^{2+}$] dependence of the open probability.
\textbf{(a)} The patch-clamp experimental data (solid stars) and our model prediction (solid line) of the [Ca$^{2+}$] dependence of the open probability at [IP$_3$] = 10 $\mu$M.
\textbf{(b)} The patch-clamp experimental data (solid stars) and our model prediction (solid line) of the [Ca$^{2+}$] dependence of the open probability at [IP$_3$] = 0.1 $\mu$M.
\textbf{(c)-(d)} The [Ca$^{2+}$] dependence of the mean closed duration.
\textbf{(c)} The patch-clamp experimental data (solid stars) and our model prediction (solid line) of the mean closed duration at [IP$_3$] = 10 $\mu$M.
\textbf{(d)} The patch-clamp experimental data (solid stars) and our model prediction (solid line) of the mean closed duration at [IP$_3$] = 0.1 $\mu$M.}\label{datafitting}
\end{center}
\end{figure}

The patch-clamp measurements of the open probability at high [IP$_3$] of 10 $\mu$M and low [IP$_3$] of 0.1 $\mu$M are illustrated by the solid symbols in Figures \ref{datafitting}(a)-(b) as a function of [Ca$^{2+}$], respectively. In addition, our model predictions of the [Ca$^{2+}$] dependence of the open probability at different [IP$_3$] are illustrated by the solid lines in Figures \ref{datafitting}(a)-(b). We see that our allosteric model fits the experimental data reasonably well. Our allosteric model, consistent with most previous IPR models, describes the bell-shaped dependence of the open probability on [Ca$^{2+}$].

Our model predicts that the [Ca$^{2+}$] dependence of the open probability at low [IP$_3$] is narrow and bell-shaped (Figure \ref{datafitting}(b)). With the increase of [IP$_3$], the top of the bell-shaped curve becomes flatter (Figure \ref{datafitting}(a)). This fact shows that a higher [IP$_3$] results in a wider region of [Ca$^{2+}$] to maintain a large open probability, which is consistent with the flat-topped [Ca$^{2+}$] dependence of the open probability at high [IP$_3$] predicted in \cite{mak2003spontaneous, baran2003integrated}.

\subsection*{Mean closed duration}
The patch-clamp measurements of the mean closed duration at high [IP$_3$] of 10 $\mu$M and low [IP$_3$] of 0.1 $\mu$M are illustrated by the solid symbols in Figures \ref{datafitting}(c)-(d) as a function of [Ca$^{2+}$], respectively. In addition, our model predictions of the [Ca$^{2+}$] dependence of the mean closed duration at different [IP$_3$] are illustrated by the solid curves in Figures \ref{datafitting}(c)-(d). We see that our allosteric model fits the patch-clamp data reasonably well. Our allosteric model predicts that the [Ca$^{2+}$] dependence of the mean closed duration changes steeply at low and high [Ca$^{2+}$] and is rather flat at [Ca$^{2+}$] between 1 $\mu$M and 10 $\mu$M.

\subsection*{Mean open duration}
\begin{figure}[!htb]
\begin{center}
\centerline{\includegraphics[width=0.8\textwidth]{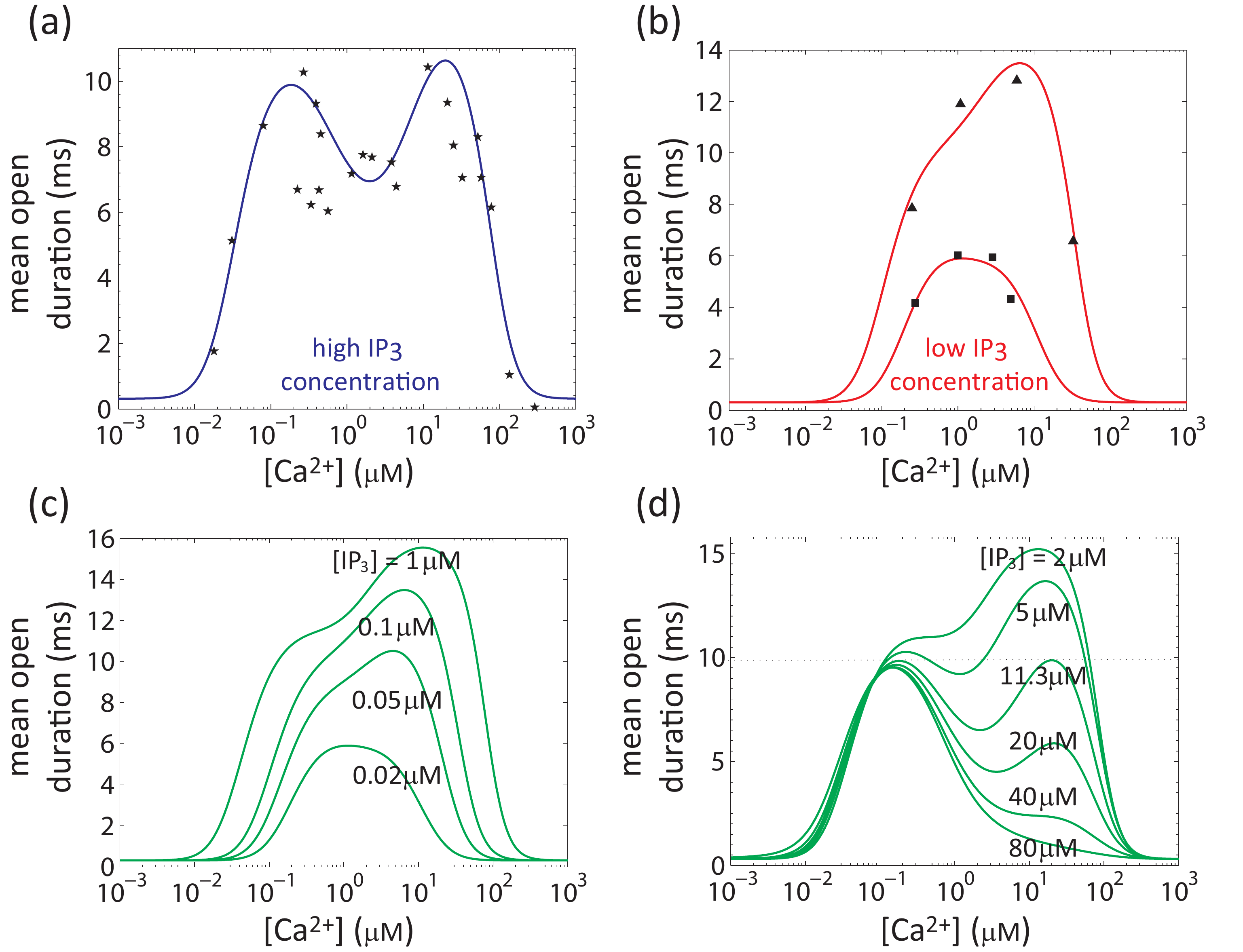}}
\caption{\textbf{(a)-(d) The [Ca$^{2+}$] dependence of the mean open duration at different [IP$_3$].}
\textbf{(a)} The patch-clamp experimental data (solid stars) and our model prediction (solid line) of the [Ca$^{2+}$] dependence of the mean open duration at [IP$_3$] = 10 $\mu$M.
\textbf{(b)} The patch-clamp experimental data and our model prediction (solid line) of the [Ca$^{2+}$] dependence of the mean open duration at [IP$_3$] = 0.02 $\mu$M and 0.1 $\mu$M. The experimental data are represented by solid squares at [IP$_3$] = 0.02 $\mu$M and are represented by solid triangles at  [IP$_3$] = 0.1 $\mu$M.
\textbf{(c)} The curves of the mean open duration versus [Ca$^{2+}$] when [IP$_3$] is lower than 2 $\mu$M. During this phase, the [Ca$^{2+}$] dependence of the mean open duration is asymmetrically bell-shaped. With the increase of [IP$_3$], the asymmetry of the curve becomes increasingly apparent.
\textbf{(d)} The curves of the mean open duration versus [Ca$^{2+}$] when [IP$_3$] is higher than 2 $\mu$M. When [IP$_3$] varies between 2 $\mu$M and 40 $\mu$M, the mean open duration is regulated by [Ca$^{2+}$] with a bimodal dependence. With the increase of [IP$_3$], the right peak decreases rapidly, but the left peak changes slightly. When [IP$_3$] is higher than 40 $\mu$M, the [Ca$^{2+}$] dependence of the mean open duration changes back to be asymmetrically bell-shaped.
}\label{tauopen}
\end{center}
\end{figure}

The patch-clamp measurements of the mean open duration at high [IP$_3$] of 10 $\mu$M and at low [IP$_3$] of 0.02 $\mu$M and 0.1 $\mu$M are illustrated by the solid symbols in Figure \ref{tauopen}(a)-(b). According to the patch-clamp data, there is an apparent bimodal [Ca$^{2+}$] dependence of the mean open duration at high [IP$_3$] of 10 $\mu$M. In addition, our model predictions of the [Ca$^{2+}$] dependence of the mean open duration at different [IP$_3$] are illustrated by the solid curves in Figures \ref{tauopen}(a)-(b). It is quite satisfactory that our allosteric model fits the patch-clamp data of the mean open duration reasonably well and much better than previous IPR models \cite{baran2003integrated, mak2003spontaneous, shuai2007kinetic, shuai2009investigation}. Particularly, we see from Figure \ref{tauopen}(a) that our model successfully describes the complicated bimodal [Ca$^{2+}$] dependence of the mean open duration at high [IP$_3$].

We see from Figures \ref{tauopen}(b) that our allosteric model also fits the patch-clamp data of the mean open duration at [IP$_3$] = 0.02 $\mu$M and 0.1 $\mu$M reasonably well. However, it is worth noting that the mean open duration at low [IP$_3$], according to both the experimental data and our model prediction, is regulated by Ca$^{2+}$ with an approximate bell-shaped dependence, instead of a bimodal dependence. Thus it is quite interesting to study how the shape of the curve of the mean open duration versus [Ca$^{2+}$] is regulated by [IP$_3$].

Our model prediction shows that with the increase of [IP$_3$], the curve of the mean open duration versus [Ca$^{2+}$] will display three different phases. When [IP$_3$] is lower than 2 $\mu$M, the [Ca$^{2+}$] dependence of the mean open duration is asymmetrically bell-shaped, as illustrated in Figure \ref{tauopen}(c). A maximal mean open duration of 15.8 ms is achieved at [IP$_3$] = 1 $\mu$M and [Ca$^{2+}$] = 10 $\mu$M. During this phase, with the increase of [IP$_3$], the asymmetry of the curve becomes increasingly apparent. At [IP$_3$] = 2 $\mu$M, the double peaks of the curve of the mean open duration versus [Ca$^{2+}$] become visible and the right peak is significantly higher than the left peak. When [IP$_3$] varies between 2 $\mu$M and 40 $\mu$M, the mean open duration becomes a bimodal function of [Ca$^{2+}$], as illustrated in Figure \ref{tauopen}(d). During this phase, the right peak decreases rapidly and the left peak changes slightly with the increase of [IP$_3$]. At [IP$_3$] = 11.3 $\mu$M, two peaks of the mean open duration has the same height of 9.9 ms. At [IP$_3$] = 40 $\mu$M, the right peak almost disappears. During the third phase that [IP$_3$] is higher than 40 $\mu$M, the right peak further decreases and the curve of the mean open duration changes back to be asymmetrically bell-shaped, as illustrated in Figure \ref{tauopen}(d).

\subsection*{Distribution of the open duration}
The patch-clamp experimental data show that the open duration of the IPR channel has a bi-exponential distribution with two time constants, $T_1\approx 20$ ms and $T_2 < 4$ ms \cite{mak2003spontaneous}. The bi-exponential distribution of the open duration has been explained theoretically by Equation \eqref{biexp}. The remaining question is whether the time constants predicted by our model are consistent with those observed in patch-clamp experiments.

We easily see from Equation \eqref{biexp} that the two time constants of the open duration are $T_1 = 1/l_1$, the mean open duration of the state $R^{open}$, and $T_2 = 1/l_2$, the mean open duration of the state $T^{open}$. Based on the parameters listed in Table \ref{parameters}, the two time constants are estimated as 17 ms and 0.32 ms, which is consistent with the experimental observation of $T_1\approx 20$ ms and $T_2 < 4$ ms.

\subsection*{Adaptation}
Over the past decade, the developments of labeled flux experiments have shown that the IPR channel is a typical biological system which performs adaptation \cite{marchant1997cooperative, marchant1998rapid, adkins1999lateral, adkins2000rapid}. Although the patch-clamp experiments are excellent at understanding the steady-state behavior of the IPR channel, it is considerably more difficult to determine the dynamic behavior from the patch-clamp recordings \cite{gin2009kinetic}. Thus it is interesting to study whether the IPR channels in their native ER membrane environment will perform adaptation based on the patch-clamped data obtained from the nuclear IPR of \emph{Xenopus} oocytes.

\begin{figure}[!htb]
\begin{center}
\centerline{\includegraphics[width=0.8\textwidth]{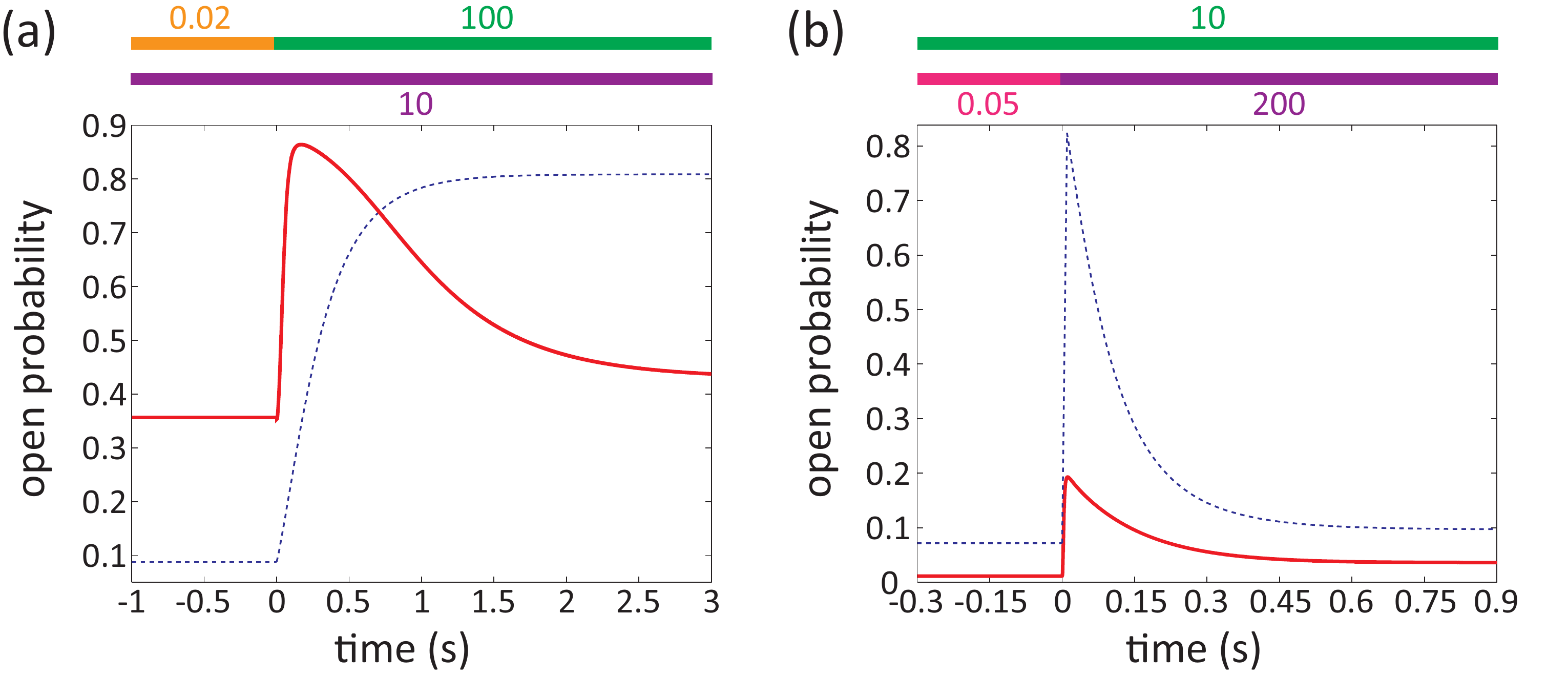}}
\caption{\textbf{(a)-(b) Adaptation of the IPR channel.}
\textbf{(a)} Dynamic responses of the IPR channel to a step increase of [IP$_3$] predicted by our model (solid line) and by the SPFMP model (dash line). In our simulation, [IP$_3$] is elevated from 0.04 $\mu$M to 100 $\mu$M at a particular time and [Ca$^{2+}$] is maintained at 10 $\mu$M, indicated by color bars above the response curves. The simulation result shows that adaptation of the IPR channel in response to step increases of [IP$_3$] is well described by our model and cannot be correctly described by the SPFMP model.
\textbf{(b)} Dynamic responses of the IPR channel to a step increase of [Ca$^{2+}$] predicted by our model (solid line) and by the SPFMP model (dash line). In our simulation, [Ca$^{2+}$] is elevated from 0.05 $\mu$M to 200 $\mu$M at a particular time and [IP$_3$] is maintained at 10 $\mu$M, indicated by color bars above the response curves. The simulation result shows that both our model and the SPFMP model can correctly describe adaptation of the IPR channel in response to step increases of [Ca$^{2+}$].}\label{adaptation}
\end{center}
\end{figure}

We first study the dynamic responses of the IPR channel to step increases of [IP$_3$]. The response of the channel open probability predicted by our model and by the SPFMP model are illustrated by the solid and dash lines in Figure \ref{adaptation}(a), respectively. In our simulation, [IP$_3$] is elevated from 0.04 $\mu$M to an ultrahigh concentration of 100 $\mu$M at a particular time and [Ca$^{2+}$] is maintained at 10 $\mu$M. According to our model prediction, the IPR channel will perform adaptation in response to a step increase of [IP$_3$], which is consistent with the observations in the labeled flux experiments. However, the SPFMP model, which assumes the detailed balance condition, cannot correctly describe the observed dynamic behavior of adaptation.

In order to gain a deeper insight into adaptation, we define two characteristic times, the reaction time and the relaxation time. The reaction time is defined as the time spent for the channel to increase from the initial open probability to the peak open probability and the relaxation time is defined as the half-life of the exponential decay from the peak open probability to the steady-state open probability. Under the above concentration conditions, our model predicts that the reaction time is between 150 ms and 200 ms and the relaxation time is between 0.5s and 1 s (Figure \ref{adaptation}(a)). Both the reaction and relaxation times predicted by our model coincide those observed in the labeled flux experiments (see, for example, Figure 1(a) in \cite{marchant1998rapid}).

We next study the dynamic responses of the IPR channel to step increases of [Ca$^{2+}$]. The response of the channel open probability predicted by our model and by the SPFMP model are illustrated by the solid and dash lines in Figure \ref{adaptation}(b), respectively. In our simulation, [Ca$^{2+}$] is elevated from 0.05 $\mu$M to 200 $\mu$M at a particular time and [IP$_3$] is maintained at 10 $\mu$M. The simulation result shows that both our model and the SPFMP model can correctly describe adaptation of the IPR channel observed in the labeled flux experiments in response to a step increase of [Ca$^{2+}$]. Our model predicts that both the reaction and relaxation times in response to a step increase of [Ca$^{2+}$] (Figure \ref{adaptation}(b)) are shorter than those in response to a step increase of [IP$_3$] (Figure \ref{adaptation}(a)).

\subsection*{Breakdown of detailed balance in the subunit model}
Recent studies on the physical essence of adaptation have made significant progresses and shown that adaptation is a nonequilibrium dynamic behavior with energy consumption \cite{lan2012energy}. In this section, we shall use the patch-clamp experimental data obtained from the nuclear IPR of \emph{Xenopus} oocytes to validate this important biophysical fact by showing that both the $R$ and $T$ subunits break detailed balance.

In order to check whether the $R$ and $T$ subunits satisfy the detailed balance condition, we introduce the following two quantities, $\gamma_R = a_1a_2b_3b_4/b_1b_2a_3a_4$ and $\gamma_T = c_1c_2d_3d_4/d_1d_2c_3c_4$. According to the Wegscheider's identity in elementary chemistry or the Kolmogorov's criterion in the theory of Markov chains, whether the $R$ subunit ($T$ subunit) satisfies the detailed balance condition depends on whether $\gamma_R = 1$ ($\gamma_T = 1$). According to the parameters listed in Table \ref{parameters}, these two quantities are estimated as $\gamma_R = 3.8\times 10^{-6}$ and $\gamma_T = 3.4\times 10^{-5}$, both of which are far less than 1. This clearly shows that both the $R$ and $T$ subunits break detailed balance to a remarkable extent, which further implies that the IPR channel is indeed a nonequilibrium system which consumes energy.

\subsection*{The energy source of the IPR channel}
Recent studies have shown that the energy sources in adaptation systems are usually energy-bearing biomolecules such as ATP, GTP and SAM \cite{lan2012energy}. For example, the osmotic stress adaptation in yeast \cite{hohmann2002osmotic}, the odor adaptation in mammalian olfactory receptors \cite{menini1999calcium, matthews2003calcium}, and the light adaptation in mammalian retinal rods \cite{nakatani1991light} are fueled by hydrolysis of ATP accompanying various phosphorylation-dephosphorylation cycles, whereas the chemotaxis of bacteria \cite{hazelbauer2008bacterial} is driven by hydrolysis of SAM accompanying methylation-demethylation cycles.

As a highly allosteric protein, the IPR channel is regulated by several heterotropic ligands including its primary ligands, IP$_3$ and Ca$^{2+}$, and other ligands such as ATP, H$^+$, and interacting proteins, as well as by redox and phosphorylation status \cite{foskett2007inositol}. The IPR channel can be phosphorylated by protein kinase A (PKA) \cite{wagner2003phosphorylation, wagner2004functional} and the binding of ATP to the channel will further modulate the ability of the channel to be phosphorylated by PKA \cite{wagner2006atp}. Phosphorylation of the IPR channel will in turn modify the channel sensitivity to IP$_3$ \cite{wagner2003phosphorylation, wagner2004functional} and now it is more generally agreed that PKA phosphorylation augments the Ca$^{2+}$ release \cite{pieper2001differential}. All these experimental results show that the IPR channel is fueled by hydralysis of ATP, which is coupled with the channel to provide the needed energy through phosphorylation-dephosphorylation cycles. During this process, high grade chemical energy is transformed into low grade heat accompanied with positive entropy production rate.

\subsection*{Breakdown of detailed balance gives rise to the bimodal behavior}
A striking fact revealed by our allosteric model is that if we assume that our subunit model satisfies the detailed balance condition, then the curve of the mean open duration versus [Ca$^{2+}$] is never of the bimodal shape. To see this fact, we rewrite Equation \eqref{opentimeformula} as
\begin{equation}
\tau_o = \frac{K_1\left(\frac{K_R}{K_T}\right)^4+K_0K_2}{l_1K_1\left(\frac{K_R}{K_T}\right)^4+l_2K_0K_2} = \frac{1}{l_1}\left(1-\frac{(l_2-l_1)K_0K_2}{l_1K_1\left(\frac{K_R}{K_T}\right)^4+l_2K_0K_2}\right).
\end{equation}
This suggests that the mean open duration $\tau_o$ and the quantity $K_R/K_T$ must depend on [Ca$^{2+}$] is the same way. If we assume that the $R$ and $T$ subunits satisfy the detailed balance condition, then we must have $\gamma_R = \gamma_T = 1$. This fact, together with Equations \eqref{rateR} and \eqref{rateT}, implies that
\begin{equation}\label{fraction}
\frac{K_R}{K_T} = \alpha\frac{(C+\frac{d_5}{c_5})(C+l)}{(C+\frac{b_5}{a_5})(C+k)}.
\end{equation}
where
\begin{equation}
\alpha = \frac{a_0d_0a_1b_2a_3(c_1c_2c_3I+c_1c_2d_3)}{b_0c_0c_1d_2c_3(a_1a_2a_3I+a_1a_2b_3)},
\end{equation}
\begin{equation}
k = \frac{a_1b_2a_3I+b_1b_2a_3}{a_1a_2a_3I+a_1a_2b_3},
\end{equation}
and
\begin{equation}
l = \frac{c_1d_2c_3I+d_1d_2c_3}{c_1c_2c_3I+c_1c_2d_3}
\end{equation}
are positive constants independent of [Ca$^{2+}$]. According to Equation \eqref{fraction}, direct calculations show that $K_R/K_T$ has at most two maximum and minimum points. However, a bimodal curve has exactly three maximum and minimum points. This suggests that the bimodal [Ca$^{2+}$] dependence of the mean open duration will never occur if the subunit model satisfies the detailed balance condition. If the subunit model breaks detailed balance, however, then Equations \eqref{rateR} and \eqref{rateT} show that $K_R/K_T$ is the quotient of two quartic functions of [Ca$^{2+}$], which may lead to the bimodal behavior.

We notice that the above analysis apparently depends on the subunit model illustrated in Figure \ref{model}(b). The readers may ask whether an equilibrium IPR model may also lead to the bimodal behavior if the subunit model becomes more complicated. In fact, in order to realize the bimodal behavior in an equilibrium model, each IPR subunit must contain at least three Ca$^{2+}$ binding sites. However, as far as we know, there is no definite experimental evidence for this choice. Thus we have good reasons to believe that the breakdown of detailed balance is an important reason that gives rise to the complicated bimodal [Ca$^{2+}$] dependence of the mean open duration at high [IP$_3$].

\section*{Discussion}

\subsection*{Comparison with the SPFMP model}
In our allosteric model, we continue to use the subunit model of the SPFMP model for the $R$ and $T$ subunits. Thus we think it necessary to discuss the differences between our model and the SPFMP model in detail.

First, the SPFMP model assumes that the IPR subunit has only one configuration and our model assumes that the IPR subunit has two configurations. As a highly allosteric protein, the IPR channel is regulated by several heterotropic ligands including its primary ligands, IP$_3$ and Ca$^{2+}$, and many other ligands such as ATP, H$^+$, and interacting proteins \cite{foskett2007inositol}. In most experiments, we mainly focus on how the gating of the IPR channel is regulated by its primary ligands, IP$_3$ and Ca$^{2+}$. However, the binding affinities of IP$_3$ and Ca$^{2+}$ are strongly influenced by the binding state of all other ligands. If we abstract the binding state of all other ligands into different configurations, then it is reasonable to assume that the IPR channel has two or more configurations.

Second, the SPFMP model assumes that the four IPR subunits are independent and our models assumes that the four IPR subunits are cooperative. In our allosteric model, we use the idea of the Monod-Wyman-Changeux allosteric model to construct our model from the subunit level to the channel level. Similar to the Monod-Wyman-Changeux model, we assume that the four IPR subunits are in the same configuration at any time and assume that the state $R^a_i$ ($i=1,2,3,4$) for the configuration $R$ and its mirror state $T^a_i$ for the configuration $T$ can convert between each other.

Third, the SPFMP model assumes that the IPR channel is open when three or four subunits are activated and our model assumes that the IPR channel is potentiated when all of the four subunits are activated and must experience a conformational change before it can contribute to the channel opening. According to our simulation result, this difference is less important and will not affect the main results of our paper. However, our assumption is consistent with most of previous IPR models, such as \cite{de1992single} and \cite{sneyd2002dynamic}.

Fourth, the SPFMP model assumes that the kinetics of the IPR subunit satisfies the detailed balance condition and our model removes this assumption. Recent developments on the physical essence of adaptation shows that adaptation systems with a negative feedback mechanism are always nonequilibrium systems which break detailed balance. This shows that the IPR channel, as a typical adaptation system with a negative feedback mechanism, must be a nonequilibrium system which consumes energy. This fact is also validated in our paper by using the patch-clamp experimental data.

Due to the above four differences, the parameters estimated in our paper (Table \ref{parameters}) are very different from those estimated in the SPFMP model. However, the magnitudes of the parameters in both the two models remain the same. In addition, the SPFMP model does not produce correct dependence of the mean open duration on [Ca$^{2+}$]. However, our model successfully describes the bimodal [Ca$^{2+}$] dependence of the mean open duration at high [IP$_3$] and reveals that the breakdown of detailed balance gives rise to this complicated bimodal behavior.

\subsection*{Comparison with the UMP model}
Besides the SPFMP model, another influential IPR model in recent years is the model developed by Ullah, Mak, and Pearson (abbreviated as the UMP model) \cite{ullah2012data}, which agrees well with the patch-clamp experimental data, especially the modal gating statistics, obtained from the nuclear IPR in insect Sf9 cells. The UMP model is a data-driven minimal model constructed in order to reproduce experimental data with as few states as possible. The transition diagram of the UMP model (see Figure 1 in \cite{ullah2012data}) is closely related to our channel model. Thus we think it necessary to discuss the similarities and differences between our model and the UMP model.

First, both the UMP model and our model are cooperative models without assuming that the IPR subunits are independent and without assuming that the ligand binding sites in each subunit are sequential or independent. Different from many previous models that assumed sequential \cite{tang1996simplification, marchant1997cooperative} or independent \cite{swillens1994transient, tang1996simplification, hirose1998allosteric, fraiman2004model} ligand binding, both the two models do not make any a priori assumption on this point, which is consistent with experimental observations showing that IP$_3$ and Ca$^{2+}$ bind to the channel cooperatively but with no specific sequential requirements \cite{mak2007rapid}.

Second, the UMP model assumes that the channel is potentiated when one or two Ca$^{2+}$ binding sites and four IP$_3$ binding sites are occupied and our model assumes that the channel is potentiated when four activating Ca$^{2+}$ binding sites and four IP$_3$ binding sites are occupied. Moreover, the UMP model assumes that each receptor state may exists in three gating modes, $L$, $I$, and $H$, similar to our assumption that each subunit state can exists in two different configurations, $R$ and $T$.

Third, the UMP model is a data-driven model and our model is a structure-motivated model. This is the most important difference between these two models. The structure of the IPR channel is only an a priori weak constraint on the UMP model. However, the assumptions of our model strongly depend on the structure of the IPR channel. A data-driven model is good at obtaining a robust estimation of the model parameters, whereas the structure and function of IPR channel are better reflected in a structure-motivated model.

Fourth, the UMP model is a channel model with no specific subunit model and our model is constructed from the subunit level to the channel level. In both the UMP model and our subunit model, the kinetics of ligand binding is governed by the law of mass action. In our channel model, however, the transition between states does not obey the law of mass action (see \emph{Methods}), which is consistent with experimental observations that the transitions between receptor states are regulated by Ca$^{2+}$ in a more complex way than simple mass action kinetics \cite{sneyd2002dynamic}.

Fifth, the UMP model assumes the detailed balance condition and our model removes this assumption. The UMP model assumes detailed balance mainly for the following two reasons. First, the detailed balance condition guarantees that the theory of aggregated reversible Markov chains can be used to derive a minimal model through an iterative data-driven approach. Second, the concept of ``occupancies" used in the UMP model only makes sense under the assumption of detailed balance. Our model removes this assumption because of the dissipative nature of adaptation and we overcome the resulting mathematical complexities by using the circulation theory of Markov chains \cite{jiang2004mathematical}.

\subsection*{Further discussions on rate constants}
By carefully checking the parameters listed in Table \ref{parameters}, we see that the reason why the $R$ and $T$ subunits break detailed balance is that both the parameters $a_1$ and $c_2$ are very close to zero. We note from Figure \ref{model}(b) that $a_1$ represents the binding affinity of the IP$_3$ binding site of the $R$ subunit when the inhibitory Ca$^{2+}$ binding site is not occupied. Thus $a_1$ is very close to zero implies that IP$_3$ binding is almost forbidden before the inhibitory Ca$^{2+}$ binding site is occupied. This shows that a free $R$ subunit is seldom potentiated by binding IP$_3$ directly, but by first binding Ca$^{2+}$ in the inhibitory site, next binding IP$_3$, and finally unbinding Ca$^{2+}$ in the inhibitory site. Similarly, we note from Figure \ref{model}(b) that $c_2$ represents the binding affinity of the inhibitory Ca$^{2+}$ binding site of the $T$ subunit when the IP$_3$ binding site is occupied. Thus $c_2$ is very close to zero means that Ca$^{2+}$ binding in the inhibitory site is almost forbidden once the IP$_3$ binding site is occupied. This shows that an important effect of IP$_3$ binding is to relieve the $T$ subunit from Ca$^{2+}$ inhibition, which coincides with the experimental conclusion of Mak and coworkers \cite{mak1998inositol}. A crucial difference between the $R$ and $T$ subunits is that a free $T$ subunit can bind IP$_3$ directly to be further potentiated, whereas a free $R$ subunit cannot.

\subsection*{Strengths and deficiencies of our model}
Just as the famous saying of George Box said: all models are wrong, but some are useful. Our model is no exception. Most previous IPR models were constructed mainly to fit experimental data and to make model predictions to guide the designs of future experiments. These are part, but not all, of the aims of this paper. Another important aim of our model is to study the role that nonequilibrium effects played in the steady-state and dynamic properties of the IPR channel. Living systems are highly dissipative, consuming energy to carry out various biological functions. Recent studies show that many important biological phenomena, such as the coherence resonance in excitable systems \cite{zhang2011stochastic}, the unidirectional movement of molecular motors \cite{qian2000mathematical}, and the switching behavior in enzyme systems \cite{jia2012kinetic}, fail to occur in equilibrium systems satisfying the detailed balance condition. In this paper, we reveal that continuous energy expenditures are also needed for the IPR channel to produce the complicated bimodal [Ca$^{2+}$] dependence of the mean open duration at high [IP$_3$] and to perform adaptation in response to step increases of ligand concentrations. This is the greatest strength of our allosteric model.

However, our allosteric model has some deficiencies. The major deficiency of our model is the discrepancy with the open probability data and the mean close duration data at low [IP$_3$] of 0.1 $\mu$M and high [Ca$^{2+}$] of 32.5 $\mu$M (see Figure \ref{datafitting}(b),(d)). Our model predicts that the open probability at low [IP$_3$] is regulated by Ca$^{2+}$ with a bell-shaped [Ca$^{2+}$] dependence, whereas the open probability data at low [IP$_3$] fail to show a downward trend at high [Ca$^{2+}$]. We do not find an appropriate explanation for this discrepancy, which also exists in previous IPR models \cite{mak2003spontaneous, baran2003integrated, shuai2007kinetic, shuai2009investigation}. We hope that repeated experiments can be made to check whether this discrepancy is due to some subtle and deep reasons or is merely due to experimental errors.

Another deficiency of our model is the lack of robustness of the estimation of the model parameters. According to our simulation, only the equilibrium constants, $a_i/b_i$ and $k_i/l_i$, and two additional parameters, $l_1$ and $l_2$, can be robustly estimated by fitting the open probability data and the mean open duration data. If more experimental data are used, as in the UMP model, then we believe that we can obtain a robust estimation of all model parameters.

In this paper, we use our allosteric model to account for some important steady-state and dynamic properties of the IPR channel, such as the bell-shaped [Ca$^{2+}$] dependence of the open probability, the bimodal [Ca$^{2+}$] the mean open duration at high [IP$_3$], the bi-exponential distribution of the open duration, and adaptation in response to step increases of ligand concentrations. We hope that our model can be further applied to account for more complicated behavior of the IPR channel such as the modal gating behavior at various ligand concentrations and the latency distributions in response to rapid changes of ligand concentrations, as has been done in the UMP model based on the patch-clamp experimental data obtained from the nuclear IPR in insect Sf9 cells. In addition, we hope that our model can be used to simulate the local concerted Ca$^{2+}$ release by clusters of IPR channels (Ca$^{2+}$ puffs) and the global propagating IPR-mediated Ca$^{2+}$ signals (Ca$^{2+}$ waves) generated through CICR \cite{berridge1997elementary}. Finally, we hope that the nonequilibrium techniques presented in this paper can be applied to study other receptor systems and single-molecule dynamics.

\section*{Methods}
In this section, we shall derive the expressions of the transition rate $a$ from the inactivated state $R^i$ to the activated state $R^a$ and the transition rate $b$ from the activated state $R^a$ to the inactivated state $R^i$ in our subunit model. Recall that the inactivated state $R^i$ is the collection of the eight states, $R_1,\cdots,R_8$. Let $X_t$ denote the state of the $R$ subunit at time $t$. Then $X_t$ is a Markov chain with state space $\{R_1,\cdots,R_8, R^a\}$. According to the probability definition of the transition rates, we have
\begin{equation}
bdt = P(X_{t+dt}=R^i|X_t=R^a) = P(X_{t+dt}=R_6|X_t=R^a) = b_0dt,
\end{equation}
where $dt$ is an infinitesimal time interval and $P(A|B)$ is the probability of the event $A$ conditional on the occurrence of the event $B$. Thus we obtain that $b = b_0$. Similarly, we have
\begin{equation}
\begin{split}
adt &= P(X_{t+dt}=R^a|X_t=R^i) \\
&= \frac{P(X_{t+dt}=R^a,X_t=R^i)}{P(X_t=R^i)} \\
&= \frac{P(X_{t+dt}=R^a|X_t=R_6)P(X_t=R_6)}{1-P(X_t=R^a)} \\
&= \frac{a_0\mu(R_6)dt}{1-\mu(R^a)},
\end{split}
\end{equation}
where $\mu(R_6)$ and $\mu(R^a)$ are the steady-state probabilities of the states $R_6$ and $R^a$, respectively. Thus we obtain that
\begin{equation}\label{a}
a = a_0\frac{\mu(R_6)}{1-\mu(R^a)}.
\end{equation}
We next consider a Markov chain $\bar{X_t}$ with state space $\{R_1,\cdots,R_8\}$ obtained from $X_t$ by deleting the state $R^a$. Let $\mu_{\bar{X}}(R_6)$ denote the steady-state probability of the state $R_6$ of the Markov chain $\bar{X_t}$. Then we easily see that
\begin{equation}\label{mubarX}
\mu_{\bar{X}}(R_6) = \frac{\mu(R_6)}{1-\mu(R_a)}.
\end{equation}
Mathematically, we can further represent the Markov chain $\bar{X}_t$ as the coupling of two Markov chains $Y_t$ and $Z_t$, where $Y_t$ describes whether $\bar{X}_t$ is in the front layer $\{R_1,R_2,R_3,R_4\}$ or the back layer $\{R_5,R_6,R_7,R_8\}$ of the subunit model (Figure \ref{model}(b)) and $Z_t$ describes whether $\bar{X}_t$ is in the lower-left corner $\{R_1,R_5\}$, the upper-left corner $\{R_2,R_6\}$, the upper-right corner $\{R_3,R_7\}$, or the lower-right corner $\{R_4,R_8\}$ of the subunit model (Figure \ref{model}(b)). Specifically, $Y_t$ is a Markov chain with state space $\{\textrm{front},\textrm{back}\}$ and transition rate matrix
\begin{equation}
Q_Y = \begin{pmatrix}
-a_5C & a_5C \\
b_5 & -b_5
\end{pmatrix}
\end{equation}
and $Z_t$ is a Markov chain with state space $\{\textrm{lower-left}, \textrm{upper-left}, \textrm{upper-right}, \textrm{lower-right}\}$ and transition rate matrix
\begin{equation}
Q_Z = \begin{pmatrix}
-(a_1I+a_4C) & a_1I & 0 & a_4C \\
b_1 & -(b_1+a_2C) & a_2C & 0 \\
0 & b_2 & -(b_2+b_3) & b_3 \\
b_4 & 0 & a_3I & -(b_4+a_3I)
\end{pmatrix}.
\end{equation}
Moreover, we can further prove that $Y_t$ and $Z_t$ are independent. This suggests that the steady-state probability of a particular state of $\bar{X_t}$ is the product of the steady-state probabilities of the corresponding states of $Y_t$ and $Z_t$. Since the state $R_6$ is in the back layer and the upper-left corner of the subunit model (Figure \ref{model}(b)), we obtain that
\begin{equation}\label{product}
\mu_{\bar{X}}(R_6) = \mu_Y(\textrm{back})\mu_Z(\textrm{upper-left}),
\end{equation}
where $\mu_Y(\textrm{back})$ is the steady-state probability of the state ``back'' of the Markov chain $Y_t$ and $\mu_Z(\textrm{upper-left})$ is the steady-state probability of the state ``upper-left'' of the Markov chain $Z_t$. Since $Y_t$ is a two-state Markov chain, we easily see that
\begin{equation}\label{muY}
\mu_Y(\textrm{back}) = \frac{a_5C}{a_5C+b_5}.
\end{equation}
According to the circulation theory of Markov chains \cite{jiang2004mathematical} which generalizes the King-Atman method in biochemistry \cite{king1956schematic}, we can prove that
\begin{equation}\label{muZ}
\mu_Z(\textrm{upper-left}) = \frac{Q_2}{Q_1+Q_2+Q_3+Q_4},
\end{equation}
where $Q_i$ is the determinant of the matrix obtained from $Q_Z$ by deleting the $i$-th row and the $i$-th column. The specific expressions of $Q_1,Q_2,Q_3$, and $Q_4$ are given in Equation \eqref{determinant}. Combining Equations \eqref{a}, \eqref{mubarX}, \eqref{product}, \eqref{muY}, and \eqref{muZ}, we finally obtain that
\begin{equation}
a = a_0\times\frac{a_5C}{a_5C+b_5}\times\frac{Q_2}{Q_1+Q_2+Q_3+Q_4}.
\end{equation}
So far, we have expressed the parameters $a$ and $b$ in the channel model as functions of the rate constants $a_i$ and $b_i$ ($i=0,1,2,3,4,5$).

\section*{Acknowledgements}
The authors are grateful to the anonymous reviewers for their valuable comments and suggestions and are grateful to Prof. Jianwei Shuai for providing the patch-clamp experimental data used in this paper. The authors gratefully acknowledge financial supports from the NSFC 11271029 and the NSFC 11171024. The first author also acknowledges financial support from the Academic Award for Young Ph.D. Researchers granted by the Ministry of Education of China.

\setlength{\bibsep}{3pt}
\small\bibliographystyle{unsrt}
\bibliography{IPR}
\end{document}